\documentclass[11pt,nofootinbib]{revtex4}

\newcommand{\be}{\begin{equation}}
\newcommand{\ee}{\end{equation}}
\newcommand{\ba}{\begin{eqnarray}}
\newcommand{\ea}{\end{eqnarray}}
\newcommand{\eps}{\epsilon}
\newcommand{\Bk}{\ensuremath{\Sigma}}
\def\cF{{\cal F}}
\def\cE{{\cal E}}
\def\cH{{\cal H}}
\def\cG{{\cal G}}
\def\G0{{\cal G}_0}
\def\F0{{\cal F}_0}
\newcommand{\vf}{\varphi} \newcommand{\nn}{\nonumber}

\newcommand{\p}{\partial}
\newcommand{\ep}{\varepsilon} 
\newcommand{\ra}{\rightarrow}

\begin{document}

\title{Vector fields in cosmology}

\author{E.~A. Davydov}
\email{davydov@theor.jinr.ru} \affiliation{Joint Institute for
Nuclear  Research, Dubna, Moscow Region RU-141980}

\begin{abstract}
Vector fields can arise in the cosmological context in different ways, 
and we discuss both abelian and nonabelian sector. In the abelian sector vector fields of the geometrical origin (from dimensional reduction and Einstein--Eddington modification of gravity) can provide a very non-trivial dynamics, which can be expressed in terms of the effective dilaton-scalar gravity with the specific potential. In the non-abelian sector we investigate the Yang--Mills $SU(2)$ theory which admits isotropic and homogeneous configuration. Provided the non-linear dependence of the lagrangian on the invariant $F_{\mu\nu}\tilde F^{\mu\nu}$, one can obtain the inflationary regime with the exponential growth of the scale factor. The effective amplitudes of the `electric' and `magnetic' components behave like slowly varying scalars at this regime, what allows the consideration of some realistic models with non-linear terms in the Yang--Mills lagrangian.
\end{abstract}

\maketitle

\section{Introduction}

The modern challenge in cosmology is to find the mechanism for the
inflation and for the present accelerated expansion. This can be
done, for example, by introducing several new (usually scalar)
fields, sometimes with rather specific properties. The other way
is to consider the modified gravity: theories with higher order
curvature corrections, $F(R)$ gravity, non-minimal coupling,
affine theory of gravity. The numerous models
can be hardly verified with our rather modest
observational possibilities.

Yet there is another approach: to use the well-known physics
and just go beyond linear lagrangians and classical limit, which
is natural at high energies of the early universe. Up to present
observational data, the most verified theories are QED and QCD,
containing spinors and gauge fields. The existence of a scalar
Higgs field is still an open question. It is curious that in
present years of the experimental search for the Higgs particle on LHC
there is an interesting breakthrough in the understanding of the
UV sector (so called `classicalization'~\cite{Dvali}), which
actually do not require the Higgs mechanism.

Therefore there arise two main branches of investigation: to
consider relatively simple `ad hoc' models with scalar fields,
which automatically suit the observed homogeneity and isotropy of
the universe, and/or to consider the more complicated but in some
sense more physically motivated sector of vector fields. Here we want to emphasize that the best way to deal with this
numerous and complicated models is to develop a general
approach, treating vector and scalar models in the same
way. For example, vector models in cosmology has to solve the
problem of diluting of the vector component, which amplitude is
scaled out as $1/a$ with large and/or rapidly growing $a$, being
the scale factor of the universe. This usually implies the
construction of some dynamical `scalars' within the vector theory.
The corresponding vector-to-scalar effective lagrangians often
inherit a specific coupling of these scalars to metric functions.

Mention that the solution to the inflation mechanism as well as
the dark energy can reveal some new aspects of gravitating
configurations like topological defects and black holes. Therefore
in such general approach one should also consider the case of
spherical and cylindrical symmetries as well as cosmological
metrics.

We organized our brief investigation in the following way.
First, in Section II, we consider often neglected vector fields of
the geometrical origin and show that in case of imposed symmetries
there can be derived the effective lagrangians with a rather
non-trivial scalar sector. Next, is Section III we investigate a
more realistic isotropic Yang--Mills $SU(2)$ theory, which in
the appropriate limit can be dynamically treated as a scalar
theory with cosmological constant.

\section{Cosmology and the geometry}

Exploring the gravitating systems, after all, one usually consider some
simplified configurations. The common procedure is a dimensional
reduction due to the intrinsic symmetry. The most popular is a
spherical one, and in four dimensions it is rather special, since
it is unique then: \be ds_{4}^2=h_{ab}dx^a
dx^b+e^{2\gamma}\left(d\theta^2+\sin^2\theta d\vf^2\right),\quad
a,\,b=0,\,1. \ee It describes the $2D$ dilaton gravity with $\gamma$
being a dilaton field. The remaining two-dimensional metric
$h_{ab}(x^a\!,x^b)$ has negative determinant, so it can always be
transformed to the form \be d
s_{2}^2=f(t,r)\left(-dt^2+dr^2\right) \label{eq:ds2} \ee by the
appropriate coordinate transformation $(x^0,\,x^1)\ra (t,\,r)$.

Another popular reduction is axial, and in most cases cylindrical,
to avoid the extra angular dependence. It is not unique, but leads
to the appearance of the massless vector fields. The cylindrical
reduction in four dimensions provides a rather complicated
configuration with two geometric vector fields along with a
so-called $\sigma$-field~\cite{ATFcyl}. In case of three
dimensions the reduction on a one-sphere will be actually
`cylindrical', providing one geometric vector field (unless
it is deliberately chosen to be vanishing). So we will proceed with the
simplest, yet non-trivial, case of a cylindrical dimensional reduction in three dimensions.

The $3D$ metric, independent on the third coordinate $\vf$, can be
decomposed as: \be ds_{3}^2=h_{ab}dx^a
dx^b+e^{2\gamma}\left(d\vf+v_a dx^a\right)^2,\quad a,b=0,1. \ee
Here $v_a$ is a two-component Kaluza--Klein vector field and
$\gamma$ is a dilaton. The curvature decomposition is standard,
providing the Einstein--Hilbert action \be S=\int
\left(R^{(2)}-\frac{e^{2\gamma}}{2}\Omega_{ab}\Omega^{ab}\right)e^{\gamma}\sqrt{-h}\,
d^2x, \label{eq:s2_1} \ee where $\Omega_{ab}=\p_{a}v_b-\p_b v_a$.
Since the vector field is massless, one can
solve it's equation of motion: \be
\p_a\left(\sqrt{-h}e^{3\gamma}\Omega^{ab}\right)=0,\quad\mathrm{then} \quad
\Omega^{ab}=Z e^{-3\gamma}\frac{\ep^{ab}}{\sqrt{-h}},\quad Z\equiv
\mathrm{const}. \label{Omega1}\ee Therefore the effective action
will describe the dilaton gravity\be
S=\int \left(e^{\gamma}R^{(2)}-Z^2 e^{-3\gamma}\right)\sqrt{-h}\,
d^2x, \label{Omega2}\ee
where the dilaton potential mimics the cosmological term.

Yet this is not enough to construct the theory with a non-trivial dynamics. The
other type of the geometrical vector field arises in the consideration of the non-Riemannian geometry with the symmetric
connection. As was shown by Einstein, the corresponding effective
theory is a standard gravity with the massive vector field
(so-called vecton) and a specific kinetic term, called later a Born--Infeld lagrangian\footnote{Actually,
the term $\sqrt{-\det(g_{\mu\nu}+F_{\mu\nu})}$ was derived
by Einstein (who used the ideas of Eddington and Weyl), first.}. The motivation for this modification was, at first, to
derive electromagnetism from gravity, and later just to remove the
restriction of the Riemannian geometry in the GR formalism which
has no fundamental reasoning. The theory was recently generalized
on case of arbitrary dimensions~\cite{ATFee}. Again, the
three-dimensional case is the most simple, since the theory
becomes linear. It's action can be written as \be
S=\int\left(R-2\Lambda-\lambda^2\Lambda\, F_{\mu\nu}F^{\mu\nu}-m^2
A_\mu A^\mu\right)\sqrt{-g}d^3x,\label{SEE}\ee with standard
parameters $\Lambda$ and $m$ being the cosmological constant and
the field mass, while $\lambda$ is a new intrinsic parameter of
the theory.

The inflationary model with the vector potential $V(A^2)$ was considered by
Ford~\cite{Ford}. The model (\ref{SEE}) is seemed to be just a simple case of the quadratic potential. But if one allows the cylindrical dimensional reduction, the presence of both types of
geometrical vector fields (vecton and KK-field) will provide a more complicated effective potential, as we will see below.

Now let us proceed with the cylindrical reduction of the theory
(\ref{SEE}). The vecton field can be decomposed by the standard
procedure \be A=A_\mu dx^\mu=\psi \left(d\vf+v_b dx^b\right)+a_b
dx^b,\quad \mu=0,1,2;\quad b=0,1, \ee on the scalar part $\psi$
and two-dimensional vector $a_b$. The field strength then will
look like \be F=F_{\mu\nu}dx^\mu \wedge dx^\nu =
f_{ab}dx^a \wedge dx^b+d\psi\wedge \left(d\vf+v_b
dx^b\right), \ee where $f_{ab}=\p_a a_b-\p_b a_a+\psi\Omega_{ab}$,
and $d\psi=\p_a\psi\,dx^a$.

Expressing the inverse $3D$ metric $g^{\mu\nu}$ in terms of the
two-dimensional metric and Kaluza--Klein vector \be
g^{\mu\nu}=\left(
             \begin{array}{cc}
               h^{ab} & -v^a \\
               -v^a &v_a v^a + e^{-2\gamma} \\
             \end{array}
           \right),
\ee where in the above expression and in what follows the
contraction over Latin indices $a,b,\ldots$ is produced with the
two-dimensional metric $h^{ab}$, one can easily calculate:
\begin{eqnarray}
  A_\mu A^\mu &=& a_b a^b+e^{-2\gamma}\psi^2, \nn\\
  F_{\mu\nu}F^{\mu\nu} &=& f_{ab}f^{ab}+2e^{-2\gamma}\p_a\psi\p^a\psi.\label{eq:A2F2}
\end{eqnarray}

In two dimensions the antisymmetric field tensors
$f_{ab},\,\Omega_{ab}$ contain only one component. Therefore the
vector theory can be substituted by the effective scalar theory.
Indeed, taking into account dimensional reduction for the
curvature part (\ref{eq:s2_1}), the equations of motion for the
vector part of the vecton action (\ref{SEE}) read:
\begin{eqnarray}
  &&\p_a\left(\sqrt{-h}e^{\gamma}\left[e^{2\gamma}\Omega^{ab}-\lambda^2\Lambda\psi  f^{ab}\right]\right) =
  0,\nn\\
  &&\lambda^2\Lambda\p_a\left(\sqrt{-h}e^{\gamma}f^{ab}\right) =m^2 e^{\gamma}\sqrt{-h}
  a^b.
\end{eqnarray}
Since in two dimensions any antisymmetric tensor is proportional
to the permutation tensor, the solution can be found in the following
form: \be \Omega^{ab}=\omega
\,e^{-\gamma}\frac{\varepsilon^{ab}}{\sqrt{-h}},\quad f^{ab}=\phi
\,e^{-\gamma}\frac{\varepsilon^{ab}}{\sqrt{-h}}, \ee where
$\phi,~\omega$ are new dynamical scalars. After substitution one
has
\begin{eqnarray}
  &\ep^{ab}\p_a\left(e^{2\gamma}\omega-2\lambda^2\Lambda\psi\phi \right) =0,& \nn\\
  &\lambda^2\Lambda\ep^{ab}\p_a\phi = m^2e^{\gamma}\sqrt{-h}\, a^b.&
\end{eqnarray}
The first equation implies that derivatives of the expression in
brackets vanish, therefore it is constant: \be
e^{2\gamma}\omega-2\lambda^2\Lambda\psi\phi=Z\equiv
\mathrm{const}. \ee The second equation allows to express the
vecton one-form $a_b$ as \be a_b
a^b=-\frac{\lambda^4\Lambda^2}{m^4}e^{-2\gamma}\p_b\phi \p^b\phi.
\ee

Finally, one can calculate the above expressions (\ref{eq:A2F2})
and obtain the following effective lagrangian\footnote{In the
effective lagrangian one should carefully check the signs at the
kinetic and potential terms after the substitution $(\p a,
a)\ra(\phi,\p\phi)$, here the signs were changed.} in terms of the
scalar amplitudes $\psi,\,\phi$: \be L(F^2\!,\,A^2)\ra
L_{eff}\left(2e^{-2\gamma}\left[(\p\psi)^2+\phi^2\right],\,
e^{-2\gamma}\left[\psi^2+\left(\p\phi\right)^2\lambda^4\Lambda^2/m^4\right]\right).
\ee In the curvature part we again can substitute the
$\Omega_{ab}$ term: \be R^{(3)}\ra
R^{(2)}-\omega^2=R^{(2)}-e^{-4\gamma}\left(Z+\lambda^2\Lambda\psi\phi\right)^2.
\ee Mention that a single KK field
 provided a dilatonic potential in the effective action~(\ref{Omega2}). Now the
corresponding potential describes the interaction of the
vecton components. And the natural massiveness of the vecton does
not allow to integrate them out as the KK field.

Collecting all terms we can write out the vecton
lagrangian after cylindrical reduction as the $2D$ dilaton gravity
with two interacting massive scalar fields: \be
L_{eff}\sqrt{-h}=\left[e^{\gamma}R^{(2)}-
e^{-\gamma}\left((\p\phi)^2/m^2+
(\p\psi)^2\right)-V_{eff}\right]\sqrt{-h}, \ee where the effective
potential is \be V_{eff}=2\Lambda e^{\gamma}+
\left[\lambda^2\Lambda\phi^2+m^2\psi^2+e^{-2\gamma}(Z+\psi\phi)^2\right]e^{-\gamma},\ee
and we rescaled $\lambda^2\Lambda\psi\ra \psi,\:
m\ra\lambda^2\Lambda m$ for convenience.

The effective potential on $(\psi,\,\phi)$ plane demonstrates the fourth
order growth accept the valleys $\psi\phi=0$, when there is a
quadratic growth or decrease, because the signs of $m^2$ and
$\lambda^2\Lambda$ are not strictly defined in the theory
(\ref{SEE}). Thus $(0,0)$ can be a minimum, local maximum or a
saddle point of the potential. In last two cases there can be
another saddle points/minima, correspondingly, depending on the
sign of $Z$. The value of the potential can be positive or
negative, providing different dynamics for the cosmological solutions.

Since this is just a toy model, we do not proceed with a detailed
investigation. We just mention that even in the simplest $D=3$
theory with vector fields coming from a geometry (cylindrical
dimensional reduction, non-Riemannian geometry) there arise a
rather non-trivial dynamics, which can be useful in
cosmology\footnote{Obviously the static configurations can be
considered as well, and some modifications to the solutions with
the event horizon will definitely appear.}. A bit more detailed
consideration will be given in the next section for the realistic
four-dimensional isotropic theory.

\section{Cosmology and the gauge theory}
The pure vector theory in cosmology usually has to deal with the
following problems: first, the isotropic configuration is
required, next, the conformal symmetry provides only the equation of state $p=\epsilon/3$.

Although it is a great task to consider the dynamical
isotropisation of the initially anisotropic/inhomogeneous
configurations (some fundamental results were obtained in~\cite{Wald}), usually the definitely isotropic
configurations are considered. In a context of an abelian field it
can be the space averaging or a so-called `cosmic
triad'~\cite{Golovlev,Triad}.

In the non-abelian sector the situation is much more favorable
because the color indices can provide some additional symmetry.
For example, with the $SU(2)$ gauge symmetry one has three vector potentials
$A_\mu^a$, which in case of the FRW metrics \be ds^2= dt^2-a(t)^2
[d\chi^2+\Bk_k(d\theta^2+\sin^2\theta d\phi^2)], \ee for closed, open
or spatially flat universe ($\Bk_1=\sin  \chi,\;\Bk_{-1}=\sinh \chi,\;
\Bk_{0}=r$), allow the homogeneous and isotropic
configuration~\cite{Galtsov:1991un}
\begin{eqnarray}\label{Fform}
F&=&F^aT^a=\dot{f}\left(T_n\,dt\wedge d\chi
     + T_\theta \Bk_k\,dt \wedge d\theta
     + T_\phi \Bk_k \sin \theta \,dt \wedge d\phi \right) \nonumber \\
     &&+ \Bk_k(f^2-k)\left(T_\phi\,d\chi \wedge d\theta
     - T_\theta \sin \theta\,d\chi \wedge d\phi
     + T_n \Bk_k  \sin \theta\,d\theta \wedge d\phi \right).
\end{eqnarray}
Here the rotating $SU(2)$ generators are used: \be T_n=\tau^a
n^a/2i,\;T_\theta=\tau^a e_\theta^a/2i,\;T_\phi=\tau^a
e_\phi^a/2i,\ee where $n^a,\,e_\theta^a,\,e_\phi^a $ are spherical
unit vectors, and $\tau^a$ are Pauli matrices. This property
remains valid also for larger gauge groups containing an embedded
$SU(2)$~\cite{Moniz:1990hf,Darian:1996mb}. Indeed, in the abelian
case the anisotropy comes form the stress-energy tensor
components, proportional to $E_i E_j,\,B_i B_j$, where $E_i,~B_i$
are the `electric' and `magnetic' parts of the field tensor. But
in the nonabelian case one has to take traces which vanish for the
configuration (\ref{Fform}) given above, when $i\neq j$.

The next task is to obtain the inflationary equation of state
$p=-\epsilon$. In the framework of Einstein gravity this can
be achieved by introducing the specific vector potential $V(A^2)$~\cite{Ford}. For the gauge field one should choose another way,
preserving the gauge symmetry. The solution is to consider the
lagrangian depending not only the square of the field tensor,
$F_{\mu\nu}F^{\mu\nu}$, but also on the invariant
$F_{\mu\nu}\tilde{F}^{\mu\nu}$.

Consider the Lagrangian $L(\cF,\cG)$ depending in an arbitrary way
on the two invariants
 \be \cF=-F^a_{\mu\nu}F^{a\mu\nu}/2,\quad
\cG=-\tilde F^{a\mu\nu}F^a_{\mu\nu}/4,\quad \tilde F^{a\mu\nu}=
\frac{\eps^{\mu\nu\lambda\tau}F^a_{\lambda\tau}}{2\sqrt{-g}}.\ee
We work in the units of the gauge field scale: $1/(gM_{Pl})$,
where $g$ is a coupling constant, so all values are dimensionless.
The linear functional $S_{\cG}=\int \cG \sqrt{-g} d^4x$ does not
depend on the metric:
 \be
S_{\cG}=-\frac12\int\frac{\eps^{\mu\nu\lambda\tau}}{\sqrt{-g}}
F_{\mu\nu}F_{\lambda\tau}\sqrt{-g} d^4x=-\frac12\int
\eps^{\mu\nu\lambda\tau}  F_{\mu\nu}F_{\lambda\tau} d^4x.
 \ee
For the configuration (\ref{Fform}) the $\cG\sqrt{-g}$
term is just a total derivative, ${\dot{f}(k-f^2)}$. But in case
of the non-linear dependence of the Lagrangian on $\cG$, one has
the following stress-energy tensor~\cite{MY}:
 \be
T_{\mu\nu}= \frac{2}{\sqrt{-g}}\frac{\delta S}{\delta
g^{\mu\nu}}=2\frac{\p L}{\p \cF}\frac{\p \cF}{\p
g^{\mu\nu}}+\left(\frac{\p L}{\p \cG}\cG-L\right)g_{\mu\nu},
 \ee
where the second term looks like the variable cosmological
constant.

The introduction of the non-linear dependence on $\cG$
term can be motivated in different ways. This can be a Born--Infeld
lagrangian with the square root, which as was shown Ref.~\cite{BI}, can
 provide the equation of state $p=-\epsilon/3$, yet insufficient
for the inflation. The quadratic dependence can be a result of the
interaction with axions. The vacuum polarization \cite{Savvidy,Dunne} at the strong
field limit provides the effective lagrangian with the logarithmic
dependence on the eigenvalues of the field tensor, which can be
expressed in terms of $\cG$ and $\cF$.

Now let us consider the YM cosmology in the inflationary regime
with an arbitrary non-linear dependence of the lagrangian on the
$\cG$ term, providing the equation of state $p=-\epsilon$. We work
then in the limit $\p L/\p\cF\ll \p L/\p\cG$, and the dominating
contribution to the equation of motion will be due to the
$\cG$-dependence. For the FRW cosmology (\ref{Fform}) in the limit
$L(\cF,\cG)\approx L(\cG)$ one has the equation of motion \be a^3\frac{\p\cG}{\p\dot
f}\frac{d}{dt}\frac{\p L}{\p\cG}+ \frac{\p
L}{\p\cG}\left[\frac{d}{dt}\left(a^3\frac{\p\cG}{\p\dot
f}\right)-a^3\frac{\p\cG}{\p f}\right]=0.\ee

Since $\cG a^3$ is a total derivative, the second term vanishes,
while the first term implies just that $\dot\cG=0$, hence
$\cG=\G0=\mathrm{const}$. Therefore the approximate solution to
the YM field is given by \be f(k-f^2/3)=\G0\int a^3dt.
\label{YMsol1}\ee From the other hand, the inflationary stage
implies the exponential dependence ${a\sim \exp{(Ht)}}$,
$H=\mathrm{const}$. The curvature term $k$ can be ignored in the
limit of the large scale factor, so the solution for the field
amplitude will be \be \label{YMsol2}
\frac{f}{a}\approx-\left(\frac{\G0}{H}\right)^{1/3}\ee for the
arbitrary non-linear dependence $L(\cG)$. This dependence will
contribute only to the value of the constant $H$ via the Friedman
equation \be H^2=\frac{8\pi}{3}\left(\frac{\p L}{\p
\cG}\cG-L\right).\ee Of course, the correct inflationary picture
implies rather the slow-roll approach, which for the arbitrary
lagrangian $L(\cG)$ is beyond the scope of the paper, but for some
particular cases it was successfully produced
in~\cite{MY,Maleknejad:2011sq}.

It is very intriguing that the inflationary solution
(\ref{YMsol1},~\ref{YMsol2}) for the non-linear configuration of
the self-interacting YM field actually looks like a massless
scalar with unusual coupling to the scale factor like in the
lagrangian $L=(\p f)^2/a$. Moreover, one can introduce the
`magnetic' and `electric' components \be \cE=\frac{\dot{f}}{a
N},\quad \cH=\frac{k-f^2}{a^2}, \quad\mbox{so that}\quad
\cF=3(\cE^2-\cH^2),\quad \cG=3\cE\cH. \ee In the regime of the
exponential growth of the scale factor they are not diluting, thus
demonstrating the scalar behavior. And, as was shown in \cite{MY}
for some particular lagrangian, they are slow-rolling during the
realistic inflation. Since most calculations of the quantum
corrections are produced in the limit of a static background
fields, one can hope that the same picture will be valid for the
slowly varying amplitudes $\cE,~\cH$ even during the inflation stage. This can justify the consideration of
non-linear terms in YM lagrangians in the cosmological context.

\section{Outlook}
In this work we discussed just a few models within the wide scope
of vector fields in cosmology. Due to the space-time symmetries
arising in most practical cases, vector models can be treated, as
a matter of fact, as some scalar-dilaton theories. This should
allow to work out the universal approach to the investigation of
both vector and scalar theories. The physically motivated theories
containing vector fields, supported by the well developed methods
within the scope of scalar models, should provide, as we hope, the
answers to the open questions in cosmology, like the inflation and
present accelerated expansion.

\section*{Acknowledgments}
I thank the Organizing Committee of 8-th International Conference on Progress in Theoretical Physics for invitation and
support, and  A.T. Filippov, D.V. Gal'tsov for useful discussions. This work was supported by
RFBR grants 11-02-01371-a, 11-02-01335-a, 11-02-12232-ofi-m-2011.


\begin{thebibliography}{99}
\bibitem{Dvali}
G.~Dvali, G.~F.~Giudice, C.~Gomez and A.~Kehagias,
  %``UV-Completion by Classicalization,''
  arXiv:1010.1415 [hep-ph] (2010).

\bibitem{ATFcyl}A.~T.~Filippov, arXiv:0605276v2 [hep-th] (2006);
 in Proc. of the workshop `Supersymmetries and Quantum Symmetries, JINR, July 27-31, JINR,
 Dubna 2006.

\bibitem{ATFee}
A.~T.~Filippov, \emph{Theor. Math. Phys.} \textbf{163}(3), 753--767 (2010), arXiv:1003.0782v2 [hep-th].

\bibitem{Ford}
L.~H.~Ford, \emph{Phys. Rev.}  D\textbf{40}, 967 (1989).

\bibitem{Wald}
R.~Wald, \emph{Phys. Rev.} D\textbf{28}, 2118 (1983).

\bibitem{Golovlev}
A.~Golovnev, V.~Mukhanov and V.~Vanchurin, \emph{JCAP}
\textbf{0806}, 009 (2008).

\bibitem{Triad}
M.~C. Bento, O.~Bertolami, P.~V.~Moniz et al., \emph{Class. Quant.
Grav.} \textbf{10}, 285--298 (1993).

\bibitem{Galtsov:1991un}
D.~V.~Galtsov and M.~S.~Volkov,
  %``Yang-Mills cosmology: Cold matter for a hot universe,''
  \emph{Phys. Lett.}  B\textbf{256}, 17 (1991).


\bibitem{Moniz:1990hf}
P.~V.~Moniz and J.~M.~Mourao,
  %``Homogeneous and isotropic closed cosmologies with a gauge sector,''
  \emph{Class. Quant. Grav. } \textbf{8}, 1815 (1991).

\bibitem{Darian:1996mb}
B.~K.~Darian and H.~P.~Kunzle,
 % ``Cosmological Einstein-Yang-Mills equations,''
  \emph{J. Math. Phys.} \textbf{38}, 4696 (1997).
  
 \bibitem{MY}
 D.~V.~Gal'tsov and E.~A.~Davydov, arXiv:1112.2943[hep-th] (2011).

\bibitem{BI}
V.~V.~Dyadichev, D.~V.~Gal'tsov, A.~G.~Zorin and M.~Y.~Zotov,
  %``Non-Abelian Born-Infeld cosmology,''
 \emph{Phys. Rev.} D\textbf{65}, 084007 (2002).

\bibitem{Savvidy}
I.~A.~Batalin, S.~G.~Matinyan and G.~K.~Savvidy, \emph{Sov. J.
Nucl. Phys.} \textbf{26}, 214 (1977); \\
\emph{Yad. Fiz.}
\textbf{26}, 407 (1977).

\bibitem{Dunne}
G.~V.~Dunne,
  %``Heisenberg-Euler effective Lagrangians: Basics and extensions,''
  arXiv:0406216 [hep-th] (2004).

\bibitem{Maleknejad:2011sq}
A.~Maleknejad and M.~M.~Sheikh-Jabbari,
  %``Non-Abelian Gauge Field Inflation,''
 \emph{Phys. Rev.}  D\textbf{84}, 043515 (2011).



\end{thebibliography}
\end{document}